\documentclass[11pt]{article}

\usepackage{amsmath,amssymb,amsfonts}
\usepackage{graphicx}
\usepackage{booktabs}
\usepackage{algorithm}
\usepackage{algpseudocode}
\usepackage{geometry}
\usepackage{hyperref}
\usepackage{float}

\title{
Explainable Regime-Aware Investing
}

\author{Amine Boukardagha}
\date{Columbia University}

\begin{document}
\maketitle

\begin{abstract}
We propose an explainable regime-aware portfolio construction framework based on a strictly causal Wasserstein Hidden Markov Model (Wasserstein HMM).
The model combines rolling Gaussian HMM inference with predictive model-order selection and template-based identity tracking using the 2-Wasserstein distance between Gaussian components, allowing regime complexity to adapt dynamically while preserving stable economic interpretation. Regime probabilities are embedded into a transaction-cost-aware mean--variance optimization framework and evaluated on a diversified daily cross-asset universe.  Relative to equal-weight and SPX Buy \& Hold benchmarks, the Wasserstein HMM achieves materially higher risk-adjusted performance (Sharpe 2.18 vs.\ 1.59 and 1.18) and substantially lower maximum drawdown (-5.43\% vs.\ -14.62\% for SPX). 
During the early 2025 equity selloff (``Liberation Day''), the strategy dynamically reduced equity exposure and shifted toward defensive assets, mitigating peak-to-trough losses. Compared to a non-parametric KNN conditional-moment estimator using the same features and optimization layer, the parametric regime model produces dramatically lower turnover and smoother weight evolution. 
The results demonstrate that regime inference stability—particularly identity preservation and adaptive complexity control—is a first-order determinant of portfolio drawdown and implementation robustness in daily allocation.
\end{abstract}

\section{Introduction}

Daily portfolio construction must reconcile two conflicting objectives: 
rapid adaptation to evolving market structure and robustness to estimation noise and trading frictions. 
At high frequency, conditional mean and covariance estimates are unstable; small perturbations in inputs can generate large shifts in optimized weights. 
When allocation depends on latent regimes, instability in regime identity or model complexity can amplify this effect, producing excessive turnover and fragile portfolio behavior.\\
Regime-aware investing addresses non-stationarity by conditioning risk and return on latent market states. 
The central promise is that regimes summarize time-varying cross-asset dependence structures and enable state-contingent allocation decisions. 
However, implementing regime models in strictly causal, rolling environments introduces two operational challenges: 
(i) the appropriate number of regimes may change over time, and 
(ii) repeated re-estimation can permute latent labels, destabilizing downstream optimization and degrading interpretability.\\
This paper introduces the \textbf{Wasserstein Hidden Markov Model (Wasserstein HMM)}, 
a regime inference architecture designed explicitly for daily portfolio construction. 
The framework combines three key elements:
\begin{enumerate}
    \item \textbf{Strictly causal rolling Gaussian HMM estimation,}
    \item \textbf{Predictive model-order selection} based on one-step-ahead log-likelihood performance, allowing the number of regimes to adapt dynamically, and
    \item \textbf{Wasserstein template tracking}, where Gaussian components are mapped to persistent regime templates using the closed-form 2-Wasserstein distance between multivariate normal distributions.
\end{enumerate}
Template anchoring provides geometrically grounded identity preservation without discrete assignment optimization, 
ensuring regime continuity through time while allowing smooth structural evolution. 
This eliminates the combinatorial instability often associated with rolling latent-state models.\\
We embed regime probabilities into a transaction-cost-aware mean--variance optimization (MVO) framework and evaluate the strategy on a diversified daily cross-asset universe. 
To isolate the economic role of regime inference, we benchmark against a non-parametric K-Nearest Neighbors (KNN) conditional-moment estimator that uses identical features and the same optimization layer.\\
Empirically, regime inference quality proves to be a first-order determinant of portfolio stability. 
The Wasserstein HMM achieves a Sharpe ratio of 2.18, substantially higher than both equal-weight diversification (1.59) and SPX Buy \& Hold (1.18), while delivering a materially smaller maximum drawdown (-5.43\% vs.\ -14.62\% for SPX). \\
The improvement is particularly evident during the early 2025 equity selloff (``Liberation Day''), where the parametric regime model detects a transition toward stress dynamics, reduces equity exposure, and reallocates toward defensive assets such as USD and bonds. 
This state-contingent adjustment results in significantly milder peak-to-trough losses relative to passive benchmarks.\\
In contrast, the non-parametric KNN approach produces highly unstable conditional moment estimates at daily frequency. 
Because nearest-neighbor sets shift discontinuously across time, the resulting MVO allocations exhibit persistent and extreme turnover, leading to higher drawdowns and less stable compounding. 
The Wasserstein HMM, by imposing structured latent-state dynamics and geometrically anchored identity tracking, yields smoother weight evolution and dramatically lower turnover.\\
The contribution of this paper is therefore twofold. 
First, we introduce a geometrically grounded regime inference framework—Wasserstein HMM—with adaptive complexity control suitable for high-frequency deployment. 
Second, we demonstrate that improvements in regime identity stability propagate directly into economically meaningful outcomes: lower turnover, reduced drawdowns, and smoother capital growth.\\
Taken together, the results show that regime-aware investing is not merely a question of whether regimes exist, but how they are inferred and stabilized. 
In daily cross-asset allocation, structured latent-state models with predictive complexity control and Wasserstein identity anchoring provide materially superior implementation characteristics relative to local similarity methods.
\section{Literature Review}
\label{sec:literature}

Modeling financial markets as evolving through latent regimes has a long tradition in econometrics and asset pricing.
Hamilton (1989) introduced the Markov-switching paradigm, establishing a statistical foundation for regime-dependent dynamics in macroeconomic variables and financial returns \cite{Hamilton1989}.
Within this tradition, Hidden Markov Models (HMMs) provide a flexible latent-state representation with well-developed filtering and likelihood-based inference procedures \cite{Rabiner1989,CappeMoulinesRyden2005}.
These tools have motivated a large literature using regime-switching models for asset allocation, where regimes act as low-dimensional summaries of time-varying expected returns, volatilities, and cross-asset dependence.\\
A central empirical finding is that allowing for regime shifts can materially improve portfolio outcomes relative to static mean--variance allocation.
Ang and Bekaert (2002, 2004) show that regime-aware international allocation can increase investor welfare and produce economically meaningful performance differences across market states \cite{AngBekaert2002,AngBekaert2004}.
Guidolin and Timmermann (2007, 2008) extend this perspective using multivariate regime-switching specifications and document sizable allocation gains when return distributions are nonlinear, skewed, or state-dependent \cite{GuidolinTimmermann2007,GuidolinTimmermann2008}.
Together, this literature emphasizes that returns are not identically distributed through time and that conditioning allocation on latent regimes can improve risk-adjusted performance.\\
However, many early regime-switching allocation studies estimate regimes on fixed samples or update them infrequently.
When HMMs are re-estimated in rolling or expanding windows at daily frequency, two operational issues become first-order.
First, the appropriate number of regimes may vary over time as market complexity changes, making any one-time model-order choice potentially fragile.
Second, repeated re-estimation can induce \emph{label switching}, where latent state identities permute across estimation dates, undermining interpretability and destabilizing downstream portfolio decisions; label switching is a well-known phenomenon in mixture and latent-variable models \cite{Stephens2000}.
These issues matter especially in daily portfolio construction, where small changes in estimated conditional moments can lead to large changes in optimized weights.\\
Recent work by Hirsa, Xu, and Malhotra (2024) addresses these practical challenges through Robust Rolling Regime Detection (R2-RD), which combines a strictly causal expanding-window HMM with (i) one-time BIC selection of the initial regime count, (ii) an emergence-only rule allowing the number of regimes to increase but not decrease, and (iii) assignment-based label matching to preserve regime identity across re-estimation dates \cite{Hirsa2024}.
R2-RD represents an important step toward operationally stable daily regime modeling by explicitly confronting model-order selection and temporal identity in rolling estimation.\\
Our methodology departs from this rule-based regime management in two directions.
First, instead of one-time information-criterion selection, we use \emph{predictive model-order selection} based on one-step-ahead log-likelihood performance, aligning regime complexity with out-of-sample predictive value under strict causality \cite{Rabiner1989,CappeMoulinesRyden2005}.
Second, rather than discrete assignment-based label matching, we enforce persistent regime identity using \emph{template tracking} grounded in optimal transport geometry: Gaussian components are mapped to persistent templates using the 2-Wasserstein distance between Gaussian distributions.
The Wasserstein geometry of Gaussian measures and its closed-form expressions provide a natural metric structure for comparing regime distributions \cite{Takatsu2011}, and computational optimal transport offers a modern unifying framework for these tools \cite{PeyreCuturi2019,Villani2009}.
By anchoring identity to evolving templates updated via exponential smoothing, we preserve interpretability while avoiding repeated combinatorial matching and reducing sensitivity to near-ties in parameter space, which are precisely the conditions that can amplify label instability \cite{Stephens2000}.\\
Beyond regime inference, our work is connected to the literature on portfolio construction under estimation error and trading frictions.
DeMiguel, Garlappi, and Uppal (2009) highlight that naive allocation rules can outperform optimized portfolios when estimation noise is severe, underscoring the need for robust moment estimates before applying mean--variance optimization \cite{DeMiguel2009}.
We therefore pair regime-conditioned expected returns with Ledoit--Wolf covariance shrinkage to stabilize high-dimensional risk estimation \cite{LedoitWolf2004}.
Finally, because daily re-optimization can induce excessive turnover, we explicitly include transaction-cost regularization in the allocation step.
Portfolio optimization with linear transaction costs and turnover-type penalties has a well-established convex-optimization treatment \cite{LoboFazelBoyd2002}, and dynamic trading models further emphasize how costs shape optimal rebalancing and smoothing of positions over time \cite{GarleanuPedersen2013}.\\
To isolate the economic role of regime inference, we benchmark our parametric HMM-based approach against a non-parametric conditional-moment estimator based on K-nearest neighbors (KNN).
Nearest-neighbor methods are classical tools for local approximation and pattern recognition \cite{CoverHart1967}, and they are standard baselines in statistical learning for capturing nonlinear conditional structure without imposing parametric assumptions \cite{HastieTibshiraniFriedman2009}.
At daily frequency, however, small perturbations in the feature space can change neighbor sets discontinuously, which can destabilize conditional moment estimates and propagate into large weight changes under mean--variance optimization—an effect that our empirical analysis highlights in turnover and drawdown comparisons.\\
In summary, the literature supports the economic value of regime-aware allocation \cite{Hamilton1989,AngBekaert2002,AngBekaert2004,GuidolinTimmermann2007,GuidolinTimmermann2008}, while also indicating that daily rolling deployment raises model-order and identity challenges \cite{Stephens2000,Hirsa2024}.
Our contribution is to replace discrete, rule-driven regime management with predictive complexity selection and geometrically anchored Wasserstein template tracking \cite{Rabiner1989,CappeMoulinesRyden2005,Takatsu2011,PeyreCuturi2019,Villani2009}, and to show that these design choices translate into implementable portfolios under shrinkage risk estimation and explicit trading frictions \cite{LedoitWolf2004,LoboFazelBoyd2002,GarleanuPedersen2013}, relative to a standard non-parametric KNN baseline \cite{CoverHart1967,HastieTibshiraniFriedman2009}.

\section{Data and Market Universe}

We evaluate strategies on a diversified cross-asset universe using daily adjusted close prices from Yahoo Finance from 2005 to 2026.
Daily log returns are computed and used to form features and portfolio returns under a strictly causal protocol.

\begin{itemize}
    \item \textbf{Equities}: S\&P 500 Index proxy (SPX)
    \item \textbf{Fixed Income}: Broad bond proxy (BOND)
    \item \textbf{Commodities}: Gold (GOLD) and Oil (OIL)
    \item \textbf{Foreign Exchange}: U.S. Dollar proxy (USD)
\end{itemize}

\section{Feature Representation}
\label{sec:features}

At each trading day $t$, the market state is represented by a feature vector:
\[
x_t =
\begin{bmatrix}
\mathbf{r}_t \\
\boldsymbol{\sigma}_t \\
\boldsymbol{m}_t
\end{bmatrix},
\]
where $\mathbf{r}_t \in \mathbb{R}^N$ is the vector of daily log returns,
$\boldsymbol{\sigma}_t \in \mathbb{R}^N$ is a 60-day rolling volatility per asset,
and $\boldsymbol{m}_t \in \mathbb{R}^N$ is a 20-day rolling mean return per asset.
All features at time $t$ use information available up to $t-1$ (strict causality).

\section{Parametric Regime Detection: Wasserstein Hidden Markov Model}
\label{sec:parametric_regime}

We refer to our regime inference architecture as the \emph{Wasserstein Hidden Markov Model (Wasserstein HMM)} framework, where market regimes correspond to different multivariate Gaussian distributions. The model is estimated on a strictly causal expanding (rolling) window, with the number of latent states selected dynamically using a predictive model-order criterion based on one-step-ahead log-likelihood performance rather than static information criteria like BIC initialization. At each re-estimation date, Gaussian HMM components are mapped to persistent regime templates using the 2-Wasserstein distance between Gaussian distributions, ensuring temporal identity without discrete assignment optimization. Templates are updated via exponential smoothing, allowing regime identities to evolve continuously while remaining stable through time. This framework combines rolling probabilistic regime inference, predictive complexity control, and optimal-transport geometry to produce adaptive yet interpretable state tracking for daily portfolio construction. We then estimate conditional moments within the expanding window and plug them into a transaction cost-aware mean variance optimization to provide daily portfolio weights.

\subsection{Feature Representation and Strict Causality}
We retain the daily feature representation. At each trading day $t$, we construct
\[
x_t = \begin{bmatrix} r_t \\ \sigma_t \\ m_t \end{bmatrix} \in \mathbb{R}^{3N},
\]
with all features computed using information available up to $t-1$.

\subsection{Predictive Model-Order Selection}

Rather than selecting the regime count once (e.g., via BIC) or enforcing monotone regime emergence,
we select the HMM order $K_t$ periodically (e.g., weekly) using a predictive scoring criterion computed strictly from past data.
Let $\mathcal{H}_t$ denote the historical training set available up to $t-1$ and $\mathcal{V}_t\subset \mathcal{H}_t$ a recent validation slice.
For each $K\in\{K_{\min},\dots,K_{\max}\}$, fit a $K$-state Gaussian HMM on $\mathcal{H}_t$ and compute the one-step-ahead predictive log score:
\[
\text{PredLL}(K;t)=\sum_{s\in\mathcal{V}_t} \log p(x_s \mid \mathcal{F}_{s-1}, K).
\]
Select
\[
K_t = \arg\max_{K\in[K_{\min},K_{\max}]} \left\{\text{PredLL}(K;t) - \lambda_K\,\text{Complexity}(K)\right\},
\]
where $\lambda_K>0$ penalizes complexity (e.g., proportional to $K$ or to the number of free parameters).

\subsection{Daily Regime Inference via Gaussian HMM}

Given $K_t$, estimate a Gaussian HMM on $\mathcal{H}_t$:
\[
z_t \in \{1,\dots,K_t\}, \qquad x_t \mid z_t=k \sim \mathcal{N}(\mu_{t,k}, \Sigma_{t,k}).
\]
The model yields filtered regime probabilities $p_{t,k}=P(z_t=k\mid \mathcal{F}_{t-1})$.

\subsection{Template-Based Regime Identity via Wasserstein Distance}

Repeated HMM re-estimation may permute latent labels across dates.
We avoid combinatorial assignment by introducing persistent \emph{templates} representing economically stable regime identities.
Let $\Theta_g=(\mu_g,\Sigma_g)$ for $g\in\{1,\dots,G\}$ denote template parameters, also referred to as regime labels.
At date $t$, each HMM component $(\mu_{t,k},\Sigma_{t,k})$ is mapped to the nearest template using the 2-Wasserstein distance:
\[
W_2^2(\mathcal{N}(\mu_1,\Sigma_1), \mathcal{N}(\mu_2,\Sigma_2)) =
\|\mu_1-\mu_2\|_2^2 + \mathrm{Tr}\!\left(\Sigma_1+\Sigma_2 - 2(\Sigma_2^{1/2}\Sigma_1\Sigma_2^{1/2})^{1/2}\right).
\]
Map each component $k$ to a template:
\[
g(k) = \arg\min_{g\le G} \; W_2(\mathcal{N}(\mu_g,\Sigma_g), \mathcal{N}(\mu_{t,k},\Sigma_{t,k})).
\]
Aggregate template probabilities:
\[
p_{t,g} = \sum_{k:\,g(k)=g} p_{t,k}.
\]
Update templates with exponential smoothing:
\[
\mu_g \leftarrow (1-\eta)\mu_g + \eta\,\bar{\mu}_{t,g},\qquad
\Sigma_g \leftarrow (1-\eta)\Sigma_g + \eta\,\bar{\Sigma}_{t,g}.
\]

\subsection{Template-Conditional Moment Aggregation}

We compute portfolio inputs as template-probability-weighted moments:
\[
\mu_t = \sum_{g=1}^{G} p_{t,g}\,\mu_g,\qquad
\Sigma_t = \sum_{g=1}^{G} p_{t,g}\,\Sigma_g.
\]

\subsection{Transaction-Cost-Aware Mean--Variance Optimization}

Given $(\mu_t,\Sigma_t)$, weights are computed daily via:
\[
\max_{w_t} \;\; \mu_t^\top w_t - \gamma\, w_t^\top \Sigma_t w_t - \tau \|w_t-w_{t-1}\|_1
\quad \text{s.t.}\quad \mathbf{1}^\top w_t=1,\; w_t\ge 0,\; \|w_t\|_\infty \le w_{\max}.
\]

\section{Non-Parametric Regime Inference: Rolling K-Nearest Neighbors Model}
\label{sec:knn}

The KNN approach does not model latent regimes explicitly.
Instead, it approximates regime-conditional moments locally by identifying historical observations whose feature vectors are closest to the current state.\\
Let $\mathcal{N}_t$ denote the index set of the $K$ nearest neighbors of $x_t$ in the expanding history.
Conditional moments are estimated by:
\[
\mu_t = \frac{1}{K} \sum_{s \in \mathcal{N}_t} r_s,
\qquad
\Sigma_t = \widehat{\mathrm{Cov}}\left(\{r_s\}_{s \in \mathcal{N}_t}\right),
\]
with covariance estimated using Ledoit--Wolf shrinkage.\\
Given $(\mu_t,\Sigma_t)$, weights are computed daily via:
\[
\max_{w_t} \;\; \mu_t^\top w_t - \gamma\, w_t^\top \Sigma_t w_t - \tau \|w_t-w_{t-1}\|_1
\quad \text{s.t.}\quad \mathbf{1}^\top w_t=1,\; w_t\ge 0,\; \|w_t\|_\infty \le w_{\max}.
\]
\section{Pseudocodes}

\subsection{Parametric Regime Investing: Wasserstein HMM + MVO}
\label{subsec:parametric_pseudocode}

\begin{algorithm}[H]
\caption{Parametric Regime Investing}
\label{alg:parametric_templates}
\begin{algorithmic}[1]
\Require Prices $\{P_t\}_{t=1}^T$ for $N$ assets; train/test split $t_0$; feature windows $(w_\sigma,w_m)$.
\Require Candidate HMM orders $K_{\min},\dots,K_{\max}$; order selection frequency $F_K$ (e.g., weekly).
\Require Validation slice length $|\mathcal{V}|$; complexity penalty $\lambda_K$; template count $G$; smoothing rate $\eta$.
\Require Risk aversion $\gamma$, transaction cost $\tau$, max-weight $w_{\max}$.
\Ensure Daily weights $\{w_t\}$ and portfolio returns $\{r_t^p\}$; template labels $\{g_t\}$.

\State Compute daily log-returns $r_t=\log(P_t)-\log(P_{t-1})\in\mathbb{R}^N$.
\State Build features $x_t=[r_t;\sigma_t;m_t]\in\mathbb{R}^{3N}$, strictly using data up to $t-1$.
\State Initialize templates $\{\Theta_g=(\mu_g,\Sigma_g)\}_{g=1}^G$ using an initial calibration window.
\State Initialize $w_{t_0}\leftarrow$ equal-weight (or zeros).

\For{$t=t_0+1$ to $T$}
    \State Form expanding history $\mathcal{H}_t=\{1,\dots,t-1\}$.
    \If{$t$ is an order-selection date (every $F_K$ days)}
        \State Set $\mathcal{V}_t \leftarrow$ last $|\mathcal{V}|$ points of $\mathcal{H}_t$.
        \For{$K=K_{\min}$ to $K_{\max}$}
            \State Fit $K$-state Gaussian HMM on $X(\mathcal{H}_t)$.
            \State Compute predictive log score $\text{PredLL}(K;t)=\sum_{s\in\mathcal{V}_t}\log p(x_s\mid\mathcal{F}_{s-1},K)$.
            \State Set score $\leftarrow \text{PredLL}(K;t)-\lambda_K\cdot \text{Complexity}(K)$.
        \EndFor
        \State Select $K_t \leftarrow \arg\max_K \text{score}(K)$.
    \EndIf

    \State Fit $K_t$-state Gaussian HMM on $X(\mathcal{H}_t)$ and compute filtered probabilities $p_{t,k}=P(z_t=k\mid\mathcal{F}_{t-1})$.
    \State Extract component parameters $(\mu_{t,k},\Sigma_{t,k})$ for $k=1,\dots,K_t$.

    \State Map each component $k$ to nearest template $g(k)$ using Wasserstein distance $W_2(\mathcal{N}(\mu_g,\Sigma_g),\mathcal{N}(\mu_{t,k},\Sigma_{t,k}))$.
    \State Aggregate template probabilities $p_{t,g}=\sum_{k:g(k)=g}p_{t,k}$.

    \State Compute template-level component averages $(\bar\mu_{t,g},\bar\Sigma_{t,g})$ from assigned components using posterior weights.
    \State Update templates: $\mu_g\leftarrow(1-\eta)\mu_g+\eta\bar\mu_{t,g}$, $\Sigma_g\leftarrow(1-\eta)\Sigma_g+\eta\bar\Sigma_{t,g}$.

    \State Compute mixture moments: $\mu_t\leftarrow\sum_g p_{t,g}\mu_g$, $\Sigma_t\leftarrow\sum_g p_{t,g}\Sigma_g$.
    \State Solve MVO:
    \[
    \max_{w}\ \mu_t^\top w-\gamma w^\top\Sigma_t w-\tau\|w-w_{t-1}\|_1
    \ \text{s.t.}\ \mathbf{1}^\top w=1,\ w\ge0,\ \|w\|_\infty\le w_{\max}.
    \]
    \State Set $w_t\leftarrow w$ and realize $r_t^p\leftarrow w_t^\top r_t$.
    \State Record dominant template label $g_t\leftarrow \arg\max_g p_{t,g}$.
\EndFor

\end{algorithmic}
\end{algorithm}
\paragraph{Diagnostics.}
We compute daily turnover $\mathrm{TO}_t=\frac{1}{2}\|w_t-w_{t-1}\|_1$, portfolio concentration $N_{\mathrm{eff},t}=\left(\sum_{i=1}^{N}w_{t,i}^2\right)^{-1}$, and track the regime-count path $\{K_t\}$.
\subsection{Non-Parametric Regime Investing: KNN + MVO}
\label{subsec:knn_mvo_pseudocode}

% --- USE USER-PROVIDED KNN PSEUDOCODE (UNCHANGED) ---
\begin{algorithm}[H]
\caption{Non-Parametric Regime Investing}
\label{alg:knn_mvo_weekly_trigger}
\begin{algorithmic}[1]
\Require
Daily prices $\{P_t\}_{t=1}^T$ for $N$ assets; train/test split date $t_0$; feature windows $(w_{\sigma},w_m)$; lookback $L$; neighbors $K$; rebalance frequency $F{=}5$; risk aversion $\lambda$; transaction cost $\tau$; max-weight constraint $w_{\max}$; triggers $(\delta_\mu,\delta_\Sigma)$.
\Ensure
Daily OOS weights $\{w_t\}$ and portfolio returns $\{r^{p}_t\}$ on $t\in\mathcal{T}_{\text{test}}$.

\State Compute daily log returns $r_t=\log(P_t)-\log(P_{t-1}) \in \mathbb{R}^N$.
\State Split returns into training $\mathcal{T}_{\text{train}}=\{1,\dots,t_0\}$ and test $\mathcal{T}_{\text{test}}=\{t_0+1,\dots,T\}$.
\State Initialize $w_{t_0} \leftarrow \mathbf{0}$ (or equal-weight), and set $(\mu_{\text{prev}},\Sigma_{\text{prev}})\leftarrow (\varnothing,\varnothing)$.

\vspace{0.25em}
\For{$t \in \mathcal{T}_{\text{test}}$ with $t \ge t_0+L$}
    \State Form the \textbf{expanding history} $\mathcal{H}_t = \mathcal{T}_{\text{train}} \cup \{t_0+1,\dots,t-1\}$ (strictly causal).
    \State Build \textbf{daily feature matrix} $X(\mathcal{H}_t)$ from returns in $\mathcal{H}_t$:
        \Statex \hspace{1.5em} (i) raw returns $r_s$,
        \Statex \hspace{1.5em} (ii) rolling volatility $\sigma_s=\mathrm{Std}(r_{s-w_{\sigma}+1:s})$,
        \Statex \hspace{1.5em} (iii) rolling momentum $m_s=\mathrm{Mean}(r_{s-w_{m}+1:s})$,
        \Statex \hspace{1.5em} and concatenate $x_s = [r_s;\sigma_s;m_s] \in \mathbb{R}^{3N}$.
    \State Let $x_t$ be the most recent feature vector (constructed using data up to $t-1$). Let $X_{\text{train}}$ be all prior feature vectors.
    \State Fit KNN on $X_{\text{train}}$ and retrieve the indices $\mathcal{N}_t$ of the $K$ nearest neighbors of $x_t$.
    \State Map neighbor indices to \textbf{neighbor return samples} $\{r_s : s \in \mathcal{N}_t\}$.
    \State Estimate \textbf{local conditional moments}:
        \Statex \hspace{1.5em} $\mu_t \leftarrow \frac{1}{K}\sum_{s\in \mathcal{N}_t} r_s$, \quad
        \Statex \hspace{1.5em} $\Sigma_t \leftarrow \text{LedoitWolfCov}\big(\{r_s\}_{s\in\mathcal{N}_t}\big)$.

        \State Compute new weights $w_t$ by solving the \textbf{transaction-cost-aware MVO} problem:
        \[
        \max_{w\in\mathbb{R}^N}\ 
        \mu_t^\top w - \lambda\, w^\top \Sigma_t w - \tau\,\|w-w_{t-1}\|_1
        \quad \text{s.t. } \mathbf{1}^\top w = 1,\ w\ge 0,\ \|w\|_\infty \le w_{\max}.
        \]
        \State Update $(\mu_{\text{prev}},\Sigma_{\text{prev}})\leftarrow (\mu_t,\Sigma_t)$.

    \State Realize daily portfolio return $r^{p}_t \leftarrow w_t^\top r_t$.
\EndFor

\State \Return $\{w_t\}_{t\in\mathcal{T}_{\text{test}}}$ and $\{r^{p}_t\}_{t\in\mathcal{T}_{\text{test}}}$.
\end{algorithmic}
\end{algorithm}
\paragraph{Diagnostics.}
We compute daily turnover $\mathrm{TO}_t=\frac{1}{2}\|w_t-w_{t-1}\|_1$, portfolio concentration $N_{\mathrm{eff},t}=\left(\sum_{i=1}^{N}w_{t,i}^2\right)^{-1}$, and track the regime-count path $\{K_t\}$.

\section{Empirical Results and Diagnostic Analysis}
\label{sec:results}

\subsection{Out-of-Sample Performance}
The parametric regime investing strategy substantially out-
performs the KNN-based strategy in both risk-adjusted performance and drawdown control.

% =========================
% Updated Table: OOS Performance Comparison (KNN vs Commercial V2.0 Parametric)
% =========================
\begin{table}[H]
\centering
\caption{Out-of-Sample Performance Comparison}
\label{tab:performance}
\begin{tabular}{lcc}
\toprule
Metric & Non-Parametric (KNN + MVO) & Parametric (Wasserstein HMM + MVO) \\
\midrule
OOS Sharpe Ratio & 1.81 & \textbf{2.18} \\
Maximum Drawdown & -12.52\% & \textbf{-5.43\%} \\
\bottomrule
\end{tabular}
\end{table}

\begin{figure}[H]
\centering
\includegraphics[width=0.95\linewidth]{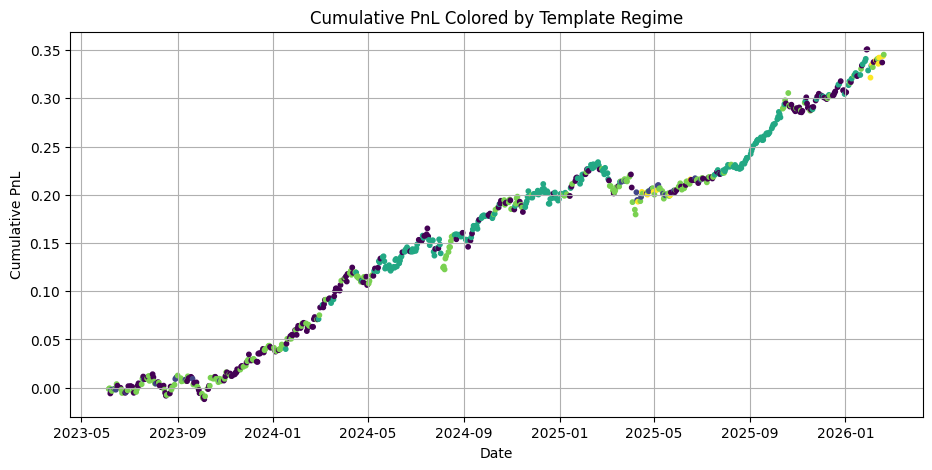}
\caption{Cumulative OOS portfolio performance for parametric regime investing.}
\label{fig:prd_pnl}
\end{figure}

\begin{figure}[H]
\centering
\includegraphics[width=0.95\linewidth]{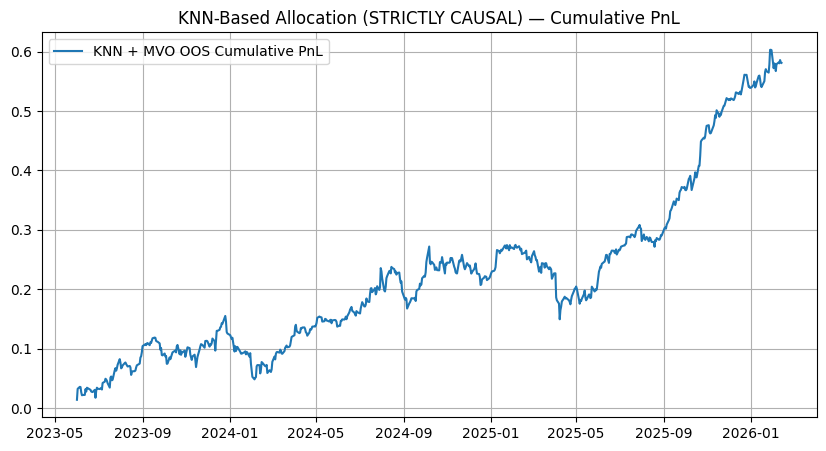}
\caption{Cumulative OOS portfolio performance for non-parametric regime investing.}
\label{fig:knn_pnl}
\end{figure}

\subsection{Rebalancing Dynamics and Turnover}

% =========================
% Updated Table: Rebalancing and Turnover Statistics (KNN vs Commercial V2.0 Parametric)
% =========================
\begin{table}[H]
\centering
\caption{Rebalancing and Turnover Statistics}
\label{tab:turnover}
\begin{tabular}{lcc}
\toprule
Metric & KNN + MVO & Parametric + MVO\\
\midrule
Average Daily Turnover & 0.5665 & \textbf{0.0079} \\
95\% Turnover Quantile & 1.0000 & \textbf{0.0504} \\
Days with $>1\%$ Turnover & 94.13\% & \textbf{14.43\%} \\
Days with $>5\%$ Turnover & 93.54\% & \textbf{5.15\%} \\
\bottomrule
\end{tabular}
\end{table}

\begin{figure}[H]
\centering
\includegraphics[width=0.95\linewidth]{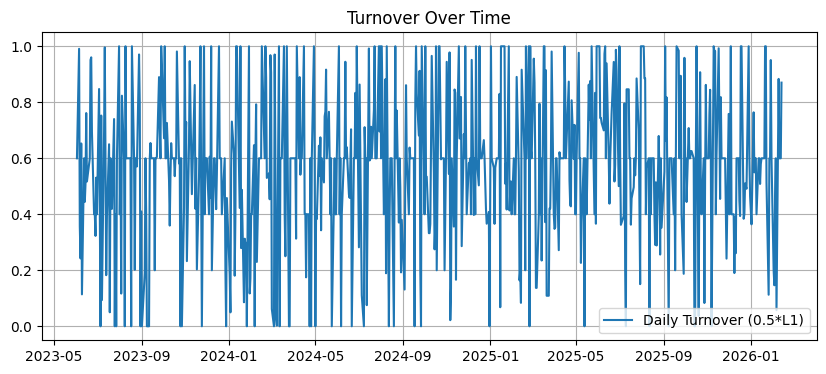}
\caption{Daily turnover time series for non-parametric regime investing.}
\label{fig:turnover_knn}
\end{figure}
The KNN strategy exhibits persistent high turnover, reflecting instability of nearest-neighbor
sets at daily frequency when used to estimate conditional moments for MVO
\begin{figure}[H]
\centering
\includegraphics[width=0.95\linewidth]{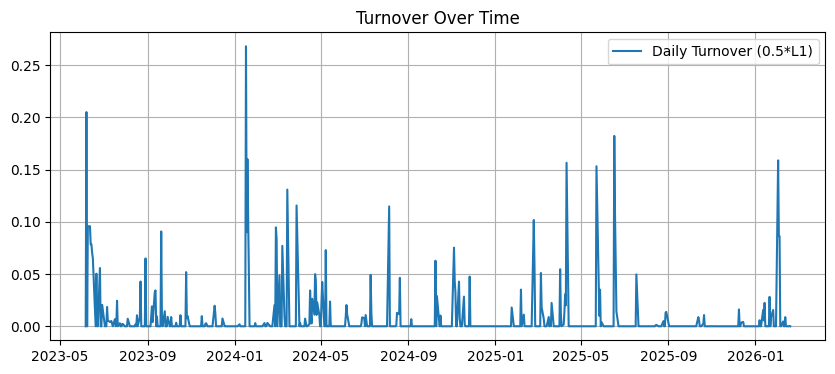}
\caption{Daily turnover time series for parametric regime investing.}
\label{fig:turnover_prd}
\end{figure}
Despite daily re-optimization, parametric regime investing produces smoother turnover dynamics due to regime persistence and more stable conditional moment estimates.
\subsection{Average Allocation and Weight Stability}

% =========================
% Updated Table: Average Allocation and Weight Dynamics (KNN vs Commercial V2.0 Parametric)
% =========================
\begin{table}[H]
\centering
\caption{Average Allocation and Weight Dynamics}
\label{tab:allocation}
\begin{tabular}{lcccc|cccc}
\toprule
& \multicolumn{4}{c}{Non-parametric regime investing} & \multicolumn{4}{c}{Parametric regime investing} \\
\cmidrule(lr){2-5} \cmidrule(lr){6-9}
Asset & Avg Wt & Wt Vol & Time $>10\%$ & Avg $|\Delta w|$
      & Avg Wt & Wt Vol & Time $>10\%$ & Avg $|\Delta w|$ \\
\midrule
SPX  & 0.221450 & 0.252030 & 0.466276 & 0.240557 & 0.256661 & 0.106745 & 0.936765 & 0.006061 \\
BOND & 0.190686 & 0.225773 & 0.461877 & 0.203325 & 0.223121 & 0.104335 & 0.954412 & 0.001619 \\
GOLD & 0.184578 & 0.244732 & 0.404692 & 0.238108 & 0.223946 & 0.041149 & 0.994118 & 0.003872 \\
OIL  & 0.201240 & 0.237979 & 0.457478 & 0.230850 & 0.004321 & 0.014390 & 0.002941 & 0.001733 \\
USD  & 0.202045 & 0.241873 & 0.456012 & 0.220146 & 0.291952 & 0.049491 & 1.000000 & 0.002543 \\
\bottomrule
\end{tabular}
\end{table}

\begin{figure}[H]
\centering
\includegraphics[width=0.98\linewidth]{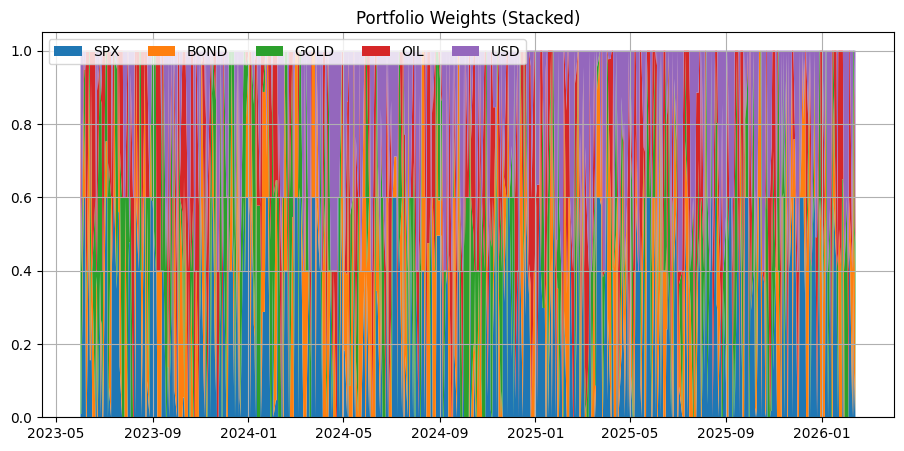}
\caption{Daily portfolio weights for non-parametric regime investing.}
\label{fig:weights_knn}
\end{figure}
The strategy exhibits frequent and large reallocations across all assets, consistent with high turnover.
\begin{figure}[H]
\centering
\includegraphics[width=0.98\linewidth]{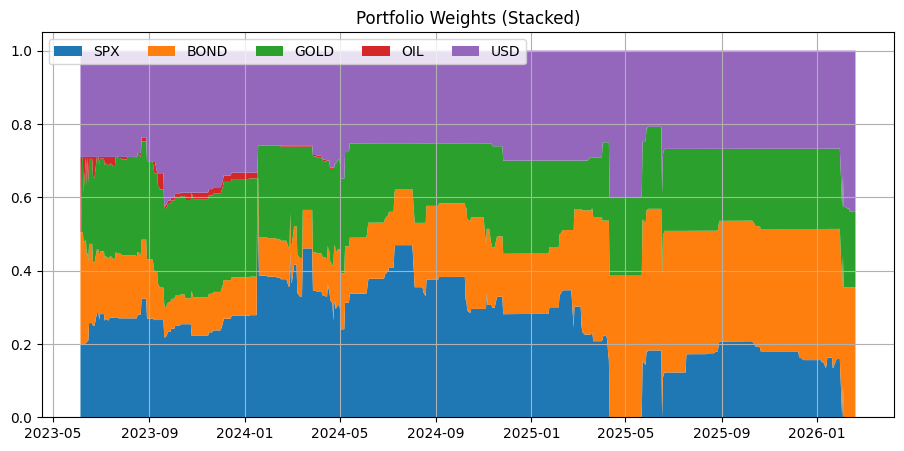}
\caption{Daily portfolio weights for parametric regime investing.}
\label{fig:weights_prd}
\end{figure}
Allocations are smoother and more persistent, with selective activation of risk assets across regimes.

\subsection{Portfolio Concentration}

% =========================
% Updated Table: Portfolio Concentration Metrics (KNN vs Commercial V2.0 Parametric)
% =========================
\begin{table}[H]
\centering
\caption{Portfolio Concentration Metrics}
\label{tab:concentration}
\begin{tabular}{lcc}
\toprule
Metric & Non-parametric regime investing & Parametric regime investing \\
\midrule
Average $N_{\text{eff}}$ & 2.07 & \textbf{3.63} \\
Median $N_{\text{eff}}$  & 1.92 & \textbf{3.70} \\
\bottomrule
\end{tabular}
\end{table}

\begin{figure}[H]
\centering
\includegraphics[width=0.95\linewidth]{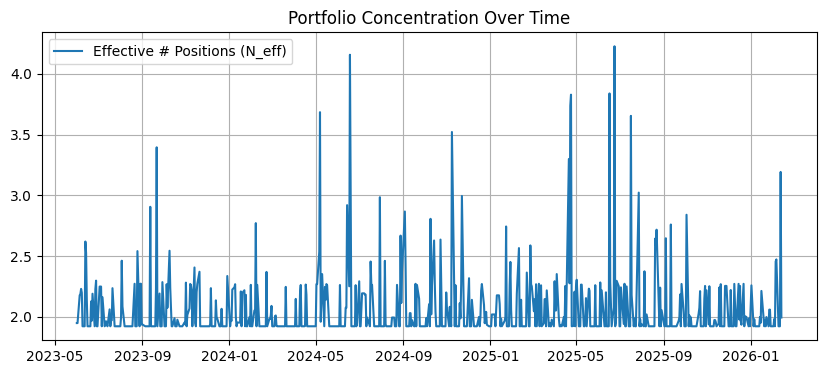}
\caption{Effective diversification $N_{\text{eff}}$ for non-parametric regime investing.}
\label{fig:neff_knn}
\end{figure}

\begin{figure}[H]
\centering
\includegraphics[width=0.95\linewidth]{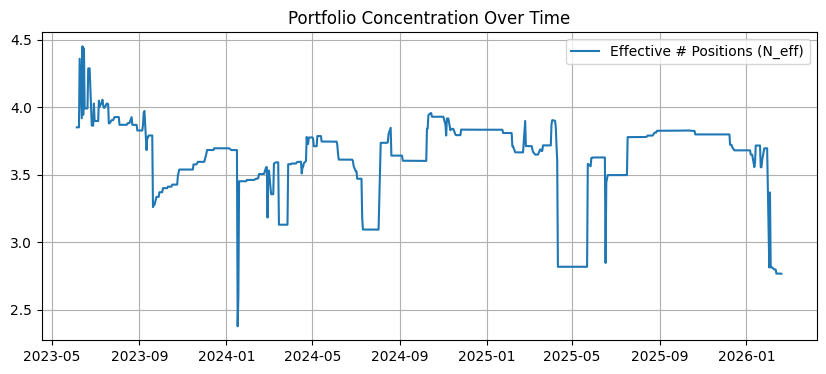}
\caption{Effective diversification $N_{\text{eff}}$ for parametric regime investing.}
\label{fig:neff_prd}
\end{figure}

\subsection{Discussion and Practical Implications}

Non-parametric KNN produces highly unstable conditional moments at daily frequency because nearest-neighbor sets vary substantially across days.
When these moments feed mean--variance optimization, the optimizer reacts aggressively, generating persistent and extreme turnover.
This is reflected in the near-universal rebalancing intensity and higher drawdowns.\\
Parametric regime detection produces stable, structured conditional moments through predictive model-order selection and template-based identity tracking using Wasserstein distance mapping between the Gaussians.\\
Wasserstein template tracking prevents label permutations and ensures smooth identity evolution without combinatorial matching.
These stability properties propagate into the MVO layer, yielding dramatically lower turnover, stronger drawdown control, and improved risk-adjusted performance.

\section{Economic Interpretation of Market Regimes}
\label{sec:template_regimes}

Because template identities persist over time, we can compute asset and portfolio performance conditional on regime. We can also interpret the regimes based on asset performance, which makes our model explainable.
In our implementation, $K$ can vary over time, but components are mapped into a fixed set of templates using Wasserstein distance mapping.\\
Empirically, one template (Template 2) receives negligible posterior mass over the evaluation window, so the realized dominant-regime set is
$\{0,1,3,4,5\}$.
For presentation, we rename these as:
\[
\text{Regime A} \equiv 0,\quad
\text{Regime B} \equiv 1,\quad
\text{Regime C} \equiv 3,\quad
\text{Regime D} \equiv 4,\quad
\text{Regime E} \equiv 5.
\]

% =========================
% Updated Table: Portfolio Performance by Regime (Templates)
% Regime mapping:
% A ≡ 0, B ≡ 1, C ≡ 3, D ≡ 4, E ≡ 5
% =========================
\begin{table}[H]
\centering
\caption{Portfolio Performance by Regime}
\label{tab:portfolio_by_regime}
\begin{tabular}{lcccccc}
\toprule
Regime & Days & Ann Mean & Ann Vol & Sharpe & Hit Rate & Max DD (within) \\
\midrule
A & 222 & 0.101301 & 0.053593 & 1.890197 & 0.558559 & -0.049624 \\
B & 29  & 0.322967 & 0.079310 & 4.072207 & 0.620690 & -0.008002 \\
C & 211 & 0.153335 & 0.061174 & 2.506528 & 0.644550 & -0.028934 \\
D & 204 & 0.096263 & 0.057581 & 1.671778 & 0.627451 & -0.038853 \\
E & 14  & 0.227562 & 0.070000 & 3.250900 & 0.642857 & -0.009568 \\
\bottomrule
\end{tabular}
\end{table}

We note that all regimes have hit rates of above 58 \% which indicates that the strategy performs very well during the backtest across all market regimes.

% =========================
% Updated Table: Asset Performance by Regime (Annualized Sharpe)
% Regime mapping:
% A ≡ 0, B ≡ 1, C ≡ 3, D ≡ 4, E ≡ 5
% =========================
\begin{table}[H]
\centering
\caption{Asset Performance by Regime (Annualized Sharpe)}
\label{tab:asset_sharpe_by_regime}
\begin{tabular}{lccccc}
\toprule
Regime & SPX & BOND & GOLD & OIL & USD \\
\midrule
A & 1.870025 & 0.695161 & 1.375891 & -0.995048 & -0.018315 \\
B & 4.227211 & 3.384970 & 2.604484 & -1.123086 & -2.747556 \\
C & 0.390222 & 2.888701 & 2.519417 & 0.425735 & -0.404131 \\
D & 0.770376 & -2.058835 & 1.268880 & 1.890725 & 2.699467 \\
E & -2.279517 & 0.807768 & 2.280656 & 0.410558 & 0.329436 \\
\bottomrule
\end{tabular}
\end{table}

\begin{figure}[H]
\centering
\includegraphics[width=0.95\linewidth]{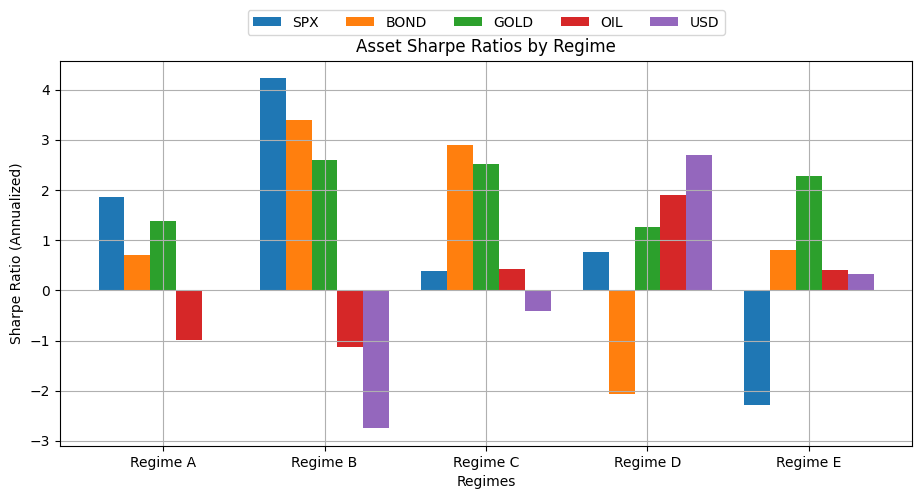}
\caption{Asset Sharpe ratios by regime.}
\label{fig:regime_sharpes}
\end{figure}

\begin{figure}[H]
\centering
\includegraphics[width=0.95\linewidth]{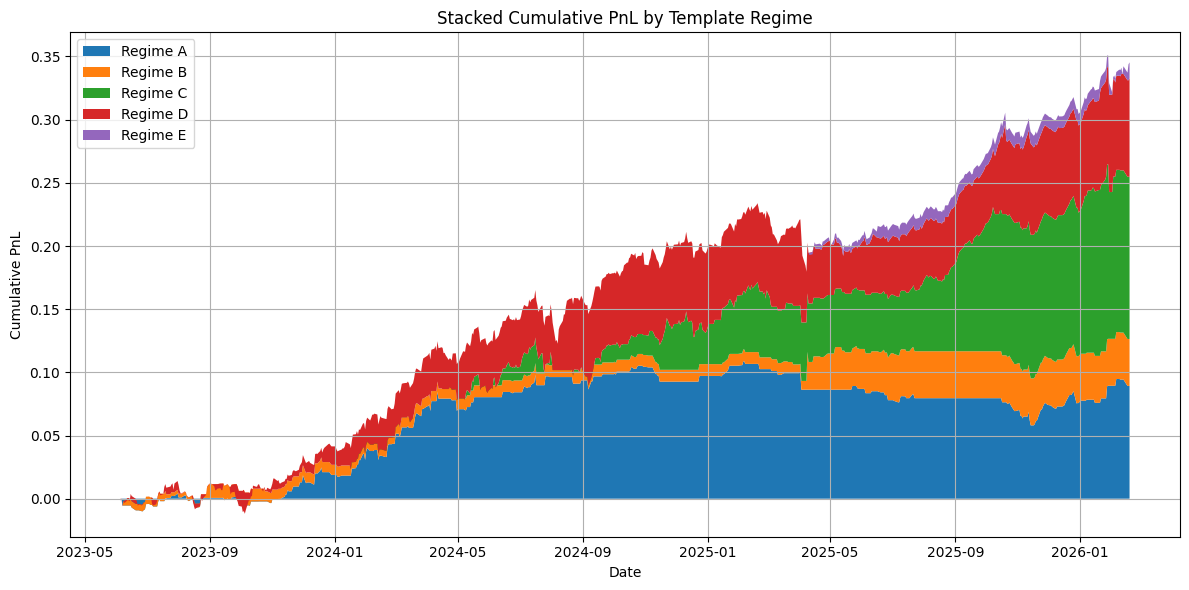}
\caption{Portfolio performance by regime (templates).}
\label{fig:pnl_by_regime}
\end{figure}

\paragraph{Regime A: Balanced Risk-On Expansion.}

Regime A exhibits strong equity performance (SPX Sharpe 1.87), positive gold performance (1.38), and supportive bond returns (0.70), while oil is negative and USD is neutral. \\
This configuration resembles a constructive macro environment with moderate growth and contained inflation. Equities and defensive diversifiers both perform well, suggesting stable risk appetite without acute macro stress. The strategy benefits from diversified exposure in this regime, reflecting broad asset participation.

\paragraph{Regime B: Broad-Based Risk Rally with Duration Support.}

Regime B is characterized by exceptionally strong Sharpe ratios for equities (4.23), bonds (3.38), and gold (2.60), while USD is sharply negative (-2.75) and oil is weak.\\
This pattern reflects a powerful risk-on regime accompanied by falling real rates and a weakening dollar. Such an environment is consistent with liquidity expansion, policy easing expectations, or synchronized global growth. The simultaneous strength of equities and bonds suggests declining discount rates and favorable financial conditions. The parametric allocation tilts toward equities and duration exposure in this state.

\paragraph{Regime C: Defensive Bond-Led Stability.}

In Regime C, bonds (2.89) and gold (2.52) lead performance, while equities remain positive but subdued (0.39). USD is modestly negative and oil is slightly positive.\\
This regime reflects a more defensive macro configuration where duration assets outperform and growth expectations moderate. It resembles a soft-landing or disinflationary stabilization phase. The strategy increases bond and gold exposure in this regime, resulting in smoother portfolio volatility.

\paragraph{Regime D: Inflationary / Dollar Strength Regime.}

Regime D displays strong oil (1.89) and USD (2.70) Sharpe ratios, positive gold (1.27), but negative bond performance (-2.06).\\
This cross-asset configuration is consistent with inflationary pressure or tightening financial conditions. Rising yields depress bond performance, while commodities and the dollar strengthen. Equities remain modestly positive but are not dominant. The regime-aware allocation reduces duration exposure and increases exposure to USD and commodities, helping insulate the portfolio from bond-driven losses.

\paragraph{Regime E: Equity Stress with Commodity Cushion.}

Regime E emerges shortly after Liberation Day. It is marked by sharply negative equity performance (SPX -2.28), while gold (2.28) and bonds (0.81) remain positive. Oil is mildly positive and USD modestly positive.\\
This regime represents an equity stress or risk-off episode. The negative equity Sharpe combined with strong gold performance suggests a flight-to-safety dynamic. Because the HMM identifies this configuration probabilistically, the allocation shifts away from equities and toward defensive assets during these episodes. \\
The Liberation Day drawdown aligns with this regime type: equities plunge while defensive assets provide cushioning. The parametric regime strategy mitigates drawdown by reallocating capital toward gold, bonds, and USD exposures consistent with the historically favorable Sharpe structure of this regime.

\subsection*{Portfolio-Level Implications}

Across regimes, the Sharpe structure reveals that:

\begin{itemize}
    \item No single asset dominates across all states.
    \item Bonds and gold provide consistent diversification benefits in multiple regimes.
    \item USD and oil performance are state-dependent and associated with inflationary or tightening environments.
    \item Equity dominance is regime-specific rather than unconditional.
\end{itemize}
The Wasserstein template framework ensures these regime identities remain stable through time, enabling economically interpretable attribution and systematic allocation shifts. Rather than relying on static diversification, the portfolio dynamically rotates toward the asset classes with historically superior risk-adjusted performance within each latent state.

\section{Benchmarking Against Passive Allocations}
\label{sec:benchmarking}

To assess whether parametric regime detection delivers economically meaningful improvements beyond passive exposure, we benchmark the strategy against two reference allocations evaluated over the identical strictly out-of-sample (OOS) period:

\begin{itemize}
    \item \textbf{SPX Buy \& Hold:} Passive exposure to the S\&P 500 index.
    \item \textbf{Equal-Weight Portfolio (20\% each):} Static allocation equally distributed across SPX, BOND, GOLD, OIL, and USD.
\end{itemize}
These benchmarks isolate:
(i) concentrated equity beta exposure, and  
(ii) naive diversification without regime conditioning.

\subsection{Risk-Adjusted Performance}

Table~\ref{tab:benchmark_performance_updated} reports OOS risk-adjusted metrics.

\begin{table}[H]
\centering
\caption{Out-of-Sample Performance Comparison}
\label{tab:benchmark_performance_updated}
\begin{tabular}{lccc}
\toprule
Strategy & Sharpe & Sortino & Max Drawdown \\
\midrule
Parametric Regime Investing & \textbf{2.18} & \textbf{2.82} & \textbf{-5.43\%} \\
Equal-Weight (20\%) & 1.59 & 2.27 & -9.87\% \\
SPX Buy \& Hold & 1.18 & 1.50 & -14.62\% \\
\bottomrule
\end{tabular}
\end{table}
While SPX Buy \& Hold ultimately achieves the highest terminal cumulative return, it does so with materially higher drawdown and volatility.
The parametric regime strategy achieves the strongest risk-adjusted performance and the smallest peak-to-trough drawdown.

\subsection{Cumulative Performance Dynamics}

\begin{figure}[H]
\centering
\includegraphics[width=0.95\linewidth]{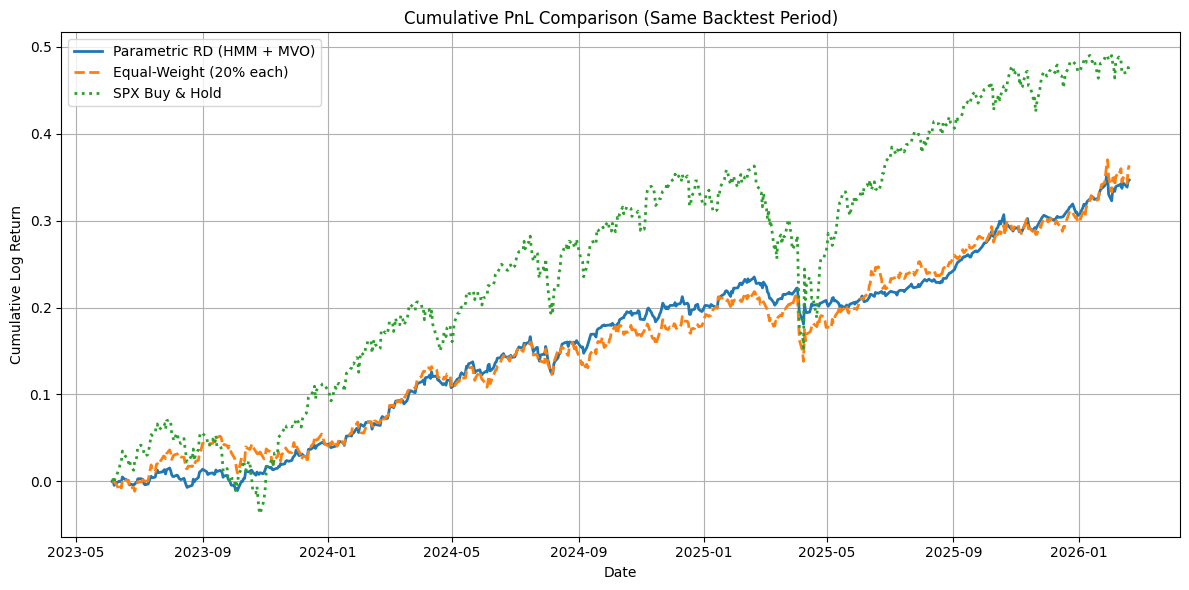}
\caption{Cumulative out-of-sample log returns: Parametric Regime Detection vs Equal-Weight vs SPX Buy \& Hold.}
\label{fig:cum_benchmark_updated}
\end{figure}

Figure~\ref{fig:cum_benchmark_updated} shows cumulative OOS log returns for all three strategies.

The SPX Buy \& Hold strategy experiences a pronounced drawdown during early 2025,
corresponding to the sharp equity selloff often referred to as the ``Liberation Day'' episode.
The equity index plunges rapidly, producing the largest interim drawdown in the sample.\\
In contrast, the parametric regime investing strategy exhibits a significantly milder drawdown during the same period.
Although it participates in equity rallies during constructive regimes,
it reduces exposure during stress regimes identified by the HMM.
The equal-weight portfolio partially mitigates concentration risk but remains mechanically exposed,
resulting in a drawdown that lies between SPX and the parametric allocation.

\subsection{Allocation Mechanism During the Shock}

The stacked allocation plot
illustrates the mechanism behind the reduced drawdown.\\
During the Liberation Day selloff:
\begin{itemize}
    \item The SPX weight contracts materially.
    \item Defensive allocations such as USD and bonds increase.
    \item Commodity exposure adjusts depending on regime-specific conditional Sharpe profiles.
\end{itemize}
This shift is not discretionary. It emerges from:
\begin{enumerate}
    \item A regime transition detected by the HMM as volatility and cross-asset covariance structures shift.
    \item Template-level conditional moment reweighting.
    \item Transaction-cost-aware MVO responding to updated risk-return trade-offs.
\end{enumerate}
The result is a systematic reduction in equity beta precisely during the stress episode.
Because regime identity is stabilized via Wasserstein template tracking,
the allocation shift occurs smoothly rather than chaotically,
and turnover remains controlled.

\subsection{Economic Interpretation}

The key insight from the benchmark comparison is not simply terminal return differences,
but \emph{path dependence and capital preservation under regime shifts}.

\begin{itemize}
    \item \textbf{SPX Buy \& Hold:} Fully exposed to equity beta at all times. Suffers the full magnitude of the Liberation Day plunge.
    \item \textbf{Equal-Weight Portfolio:} Diversifies across assets but does not adapt to regime changes. Still experiences meaningful drawdown.
    \item \textbf{Parametric Regime Investing:} Conditions allocation on latent state probabilities. Reduces equity exposure during stress regimes and reallocates toward defensive assets.
\end{itemize}
The cumulative PnL chart demonstrates visibly smoother capital compounding for the parametric strategy.
The allocation chart confirms that drawdown control is driven by endogenous state detection rather than ad hoc overlays.

\subsection{Implications for Regime-Aware Investing}

The Liberation Day episode provides a natural stress test of regime-aware allocation.
While SPX Buy \& Hold maximizes exposure to equity risk premia,
it does so at the cost of substantial interim drawdowns.
The parametric regime strategy sacrifices some extreme upside in pure equity rallies,
but it delivers:
\begin{itemize}
    \item Smaller peak-to-trough drawdowns,
    \item Higher Sharpe and Sortino ratios,
    \item Smoother weight evolution,
    \item More stable capital growth.
\end{itemize}
This evidence supports the central thesis of the paper:
regime inference quality directly affects portfolio path stability.
When regime transitions are detected probabilistically and identities are preserved through Wasserstein template anchoring,
allocation adapts in a structured and economically interpretable manner,
leading to materially improved downside control during market shocks.

\section{Conclusion}

This paper develops a daily regime-aware allocation framework built around the Wasserstein Hidden Markov Model (Wasserstein HMM), a strictly causal rolling Gaussian HMM with predictive model-order selection and template-based identity tracking using the 2-Wasserstein distance.\\ 
By combining adaptive complexity control with geometrically anchored regime identities, the framework is designed explicitly for high-frequency deployment where estimation noise and label instability can materially affect portfolio decisions.\\
Empirically, regime inference quality proves to be a first-order determinant of implementation performance. 
Relative to passive benchmarks, the Wasserstein HMM delivers higher risk-adjusted returns and substantially smaller drawdowns. 
During the early 2025 equity selloff (“Liberation Day”), the model detected a transition toward a stress regime and systematically reduced equity exposure, reallocating toward defensive assets. \\
This state-contingent adjustment resulted in significantly milder peak-to-trough losses compared to SPX Buy \& Hold and static diversification.\\
Compared to a non-parametric KNN conditional-moment estimator using identical features and optimization constraints, the parametric regime framework generates dramatically lower turnover and smoother weight evolution. \\
Local similarity methods produce discontinuous moment estimates that propagate into excessive trading, whereas structured latent-state dynamics with template anchoring stabilize conditional inputs to optimization.\\
The results suggest that in daily cross-asset allocation, the central challenge is not merely detecting regimes, but preserving stable regime identity while allowing complexity to adapt. 
Geometrically grounded template tracking within a predictive HMM framework provides a principled and economically interpretable solution.

\section{Acknowledgment}
The author would like to thank Alex Saunders of Citi research, Joerg Osterrieder of the University of Twente, Ali Hirsa of Columbia University, and Nathaniel Rupert of Breven Howard for numerous insightful conversations that helped lead to the development of this work.

\end{document}